\documentclass[twocolumn,superscriptaddress,nobalancelastpage,amsmath,10pt,pra,a4paper]{revtex4-1}

\usepackage{graphicx}
\usepackage{amsfonts,float}
\begin{document}

\title{
Heisenberg versus standard scaling in quantum metrology with Markov generated states and monitored environment}

\author{Catalin Catana}

\author{M\u{a}d\u{a}lin Gu\c{t}\u{a}}
\email{madalin.guta@nottingham.ac.uk}

\affiliation{School of Mathematical Sciences, University of Nottingham, Nottingham, NG7 2RD, UK}

%

\vspace{3mm}
\begin{abstract}  
Finding  optimal and noise robust probe states is a key  problem in quantum metrology. In this paper we propose Markov dynamics as a possible mechanism for generating such states, and show how the Heisenberg scaling emerges for systems with multiple `dynamical phases' (stationary states), and noiseless channels. We model noisy channels by coupling the Markov output to `environment' ancillas, and consider the scenario where the environment is monitored to increase the quantum Fisher information of the output. In this setup we find 
that the survival of the Heisenberg limit depends on whether the environment receives `which phase' information about the memory system. 

\end{abstract}

\maketitle

\section{Introduction}
 The accurate estimation of unknown parameters is a fundamental task in quantum technologies, with applications ranging from  spectroscopy \cite {Wineland} and  interferometry \cite{Holland,Lee}, to atomic clocks \cite{Valencia, Burgh, Buzek} and gravitational wave detectors \cite{Caves,Ligo}. In a typical metrological protocol \cite{Giovanetti&Science},  a quantum transformation $T_\theta$
 is applied (in parallel) to each component of a `probe' ensemble of $n$  quantum systems initially prepared in the joint state $|\Psi^n\rangle$. The ensemble is subsequently measured and an estimator $\hat{\theta}_n$ of $\theta$ 
is computed. While for uncorrelated states the mean square error scales as $1/n$ (standard scaling), if quantum resources such as entanglement or squeezing are used in the preparation stage, the precision can be enhanced to 
$1/n^2$ (Heisenberg scaling) if $T_\theta$ is unitary
\cite{Holland,Giovanetti&Science,vGiovannetti&Lloyd}.

However, when noise and decoherence are taken into account, they typically lead to a `downgrading' of the Heisenberg scaling to the standard one, but a `quantum enhancement' is nevertheless achievable in the form of a constant factor that increases with decreasing noise level \cite{Huelga,Banaszek&Rafal,Dorner,Guta&Rafal, Chaves&Acin}. In this setup, new tools for deriving upper bounds on the quantum Fisher information of the final state have been developed in \cite{Guta&Rafal,Escher, Zwierz,Ji}. We note also that the Heisenberg limit \emph{can} be preserved for some noise models \cite{Andre, Auzinsh} or by using quantum error correction techniques \cite{Kessler13,Arrad13,Duer13,Ozeri13} in certain modified metrological settings.

\vspace{2mm}

\begin{figure}[H]
   \begin{center}  
 \scalebox{0.5}{\includegraphics[width=0.99\textwidth]{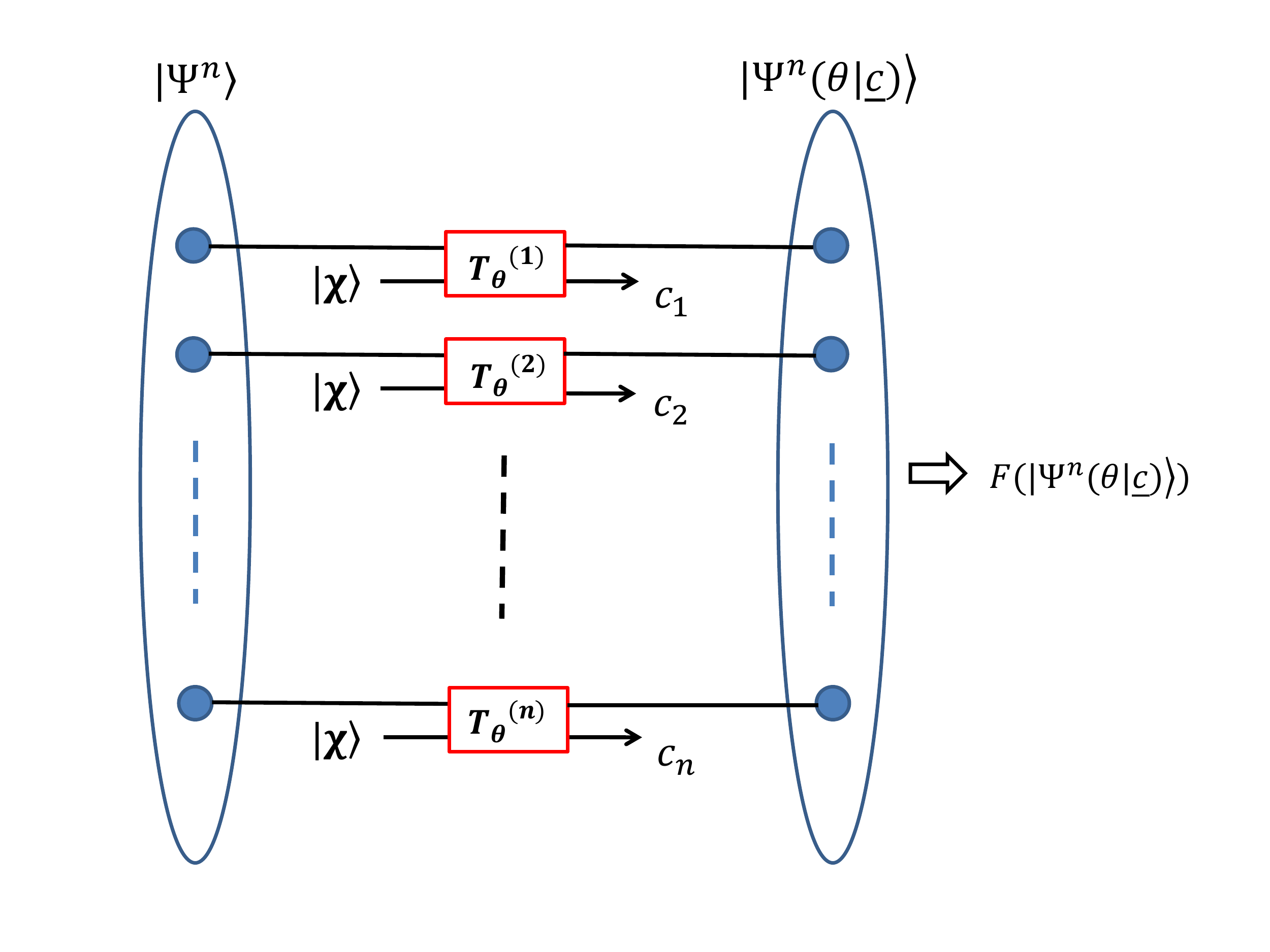}}
   \end{center}
\caption{ (Color online) Quantum metrology with $n$ noisy channels acting on a pure state input $|\Psi^n\rangle$. The channel $T^{(i)}_\theta$ is monitored by measuring the associated `environment' and the result $c_i$ is obtained. 
Conditional on the measurement record $\underline{c}:= (c_1, \dots, c_n)$, the final probe state has quantum Fisher information $F(\Psi^n(\theta|\underline{c}))$.}
\label{fig:model}
\end{figure}

The aim of this paper is to explore quantum metrology in a novel setup characterised by two key features. Firstly, we model the channel $T_\theta$ as coupling with an ancilla (environment) and we assume that the latter can be monitored by means of measurements, as illustrated in Figure \ref{fig:model}. The outcome $\underline{c}$ of the measurement provides additional information, which generally improves the estimation efficiency, and even restores the Heisenberg limit in certain models. Secondly, the n-partite probe state is generated as output of a quantum Markov chain, which is similar to the matrix product states (MPS) ansatz proposed in \cite{Jarzyna&Rafal}. More concretely, the probe systems are initially independent and identically prepared, and interact successively with a `memory' system which imprints correlations into the ensemble. This specific preparation method allows us to apply system identification techniques for quantum Markov dynamics \cite{Guta,Guta&Kiukas}, and to identify the mechanism responsible for the Heisenberg scaling and its degrading. 
In this setting, we observe that that if  the Markov transition operator used for generating the probe state is primitive (irreducible and aperiodic), then the quantum Fisher information (QFI) scales linearly with $n$ even with full access to the environment, and the standard limit holds. We therefore consider Markov models with multiple `dynamical phases' (invariant spaces), and investigate the evolution of the probe state QFI, conditional on the environment measurement record. We show that one of the following two scenarios can occur. If the  environment measurement  does not distinguish between the dynamical phases, and the memory is started in a superposition of different phases, then the QFI of the conditional output state scales as $n^2$, and the Heisenberg scaling holds. In a two-phases Markov dynamics for instance, for large $n$ the conditional memory-probe state becomes a `macroscopic' superposition 
$$
|\Psi^n(\theta|\underline{c}) \rangle= |\Psi^{n,0}(\theta|\underline{c})\rangle+|\Psi^{n,1}(\theta|\underline{c})\rangle
$$ 
of components whose weights remain constant even when the environment is observed. The quantum Fisher information of this superposition is proportional to the variance of the generator $\bar{G}$ responsible for the parameter change. 
For each component $a=0,1$, the mean value of the generator increases linearly with $n$ as $ng_a$, so that if $g_0\neq g_1$, the variance of $\bar{G}$ with respect tot the superposition, grows as $n^2(g_0-g_1)^2$.

Alternatively, if the environment receives `which phase' information about the memory, the QFI may have an initial quadratic scaling but becomes linear as the memory system is `collapsed' to one of the phases. In this case, by simulating measurement trajectories we can estimate the average conditional QFI as a function of $n$ and identify the optimal number of iterations of the Markov dynamics. In particular, this provides an upper bound to the QFI of the probe state in the absence of monitoring.

\vspace{2mm}

In section \ref{sec.monitored} we discuss the monitored environment setup, the associated notion of Fisher information, and present a toy example where monitoring restores the Heisenberg scaling. Section \ref{sec.Markov} describes the full setup including the Markov generated probe states. In section \ref{sec.example} we analyse a dephasing channel example, and show how standard or Heisenberg scaling ca be achieved depending on the chosen Markov dynamics. We finish with comments on possible further investigations.

\section{Metrology with monitored environment}\label{sec.monitored}

In this section we describe the environment monitoring scheme, and compare the associated Fisher information to that of the standard metrology setup. We refer to the appendix for a brief review of the definition and statistical interpretation of the quantum Fisher information.

\vspace{2mm}

The noisy channels $T_\theta$ are modelled by coupling each probe system unitarily to an individual ancilla representing its environment, as illustrated in Figure \ref{fig:model}. If the ancilla is initially prepared in state $|\chi\rangle$, and the interaction is described by the unitary $W_\theta$, then $T_\theta$ has Kraus decomposition 
$$
T_\theta (\rho) = \sum_c A_c^\theta \rho A_c^{\theta\dagger},\qquad A_c^\theta := \langle c | W_\theta |  \chi\rangle,
$$
where $|c\rangle$ is chosen to be the basis in which the environment is measured. By monitoring the ensemble of $n$ ancillas we obtain the outcome $\underline{c}=(c_1,\dots,c_n)$, an the conditional final state of the probe ensemble is 
$$
|\Psi^n(\theta|\underline{c}) \rangle:=  
\frac{A^\theta_{c_n}\otimes \dots \otimes A^\theta_{c_1} |\Psi^n \rangle}{\sqrt{p^n(\underline{c}|\theta) }}, 
$$
where $p^n(\underline{c}|\theta) $ is the probability of the outcome $\underline{c}$
$$
p^n(\underline{c}|\theta) :=\|A^\theta_{c_n}\otimes \dots \otimes A^\theta_{c_1} \Psi^n\|^2 . 
$$
The measurement data $\underline{c}$ is fed into the design of the  final measurement which aims to extract the maximum amount of information about  $\theta$.

Our figure of merit for estimation is the \emph{total} Fisher information of the available `data' consisting of the classical result $\underline{c}$ \emph{and} the conditional quantum state of the probe $|\Psi^n(\theta | \underline{c})\rangle$.  This can be written as (see appendix)
\begin{equation*}\label{eq.f.monitor}
F^n_{total}(\theta) = F^n_{cl} (\theta) + \sum_{\underline{c}} p^n(\underline{c}|\theta) F_q ( \Psi^n(\theta|\underline{c})) 
\end{equation*}
where the first term on the right side is the \emph{classical Fisher information} of $\underline{c}$, while the second is the 
\emph{average quantum Fisher information} of the final conditional state. By comparison, the figure of merit for the standard (no monitoring) quantum metrology setting is the quantum Fisher information $F_q(\rho^n_\theta)$ of the (average) final probe state 
$$
\rho^n_\theta= 
\sum_{\underline{c}}p^n(\underline{c}) | \Psi^n(\theta|\underline{c})\rangle \langle \Psi^n(\theta|\underline{c})|
=T_\theta^{\otimes n}(|\Psi^n\rangle\langle \Psi^n|).
$$ 
Since monitoring provides additional information, the following inequality holds
$$
F^n_{total}(\theta) \geq F_q(\rho^n_\theta).
$$
We illustrate our setup with the following toy example.  The qubit channel $T_\theta$ is the convex combination of unitary rotations
\begin{equation}\label{eq.ttheta}
T_\theta \left(\rho\right)=\sum_{j \in \{0,\dots,3\}}\lambda_j e^{i\theta\sigma_j}\rho e^{-i\theta\sigma_j} 
\end{equation}
where $\sigma_j$ are the Pauli matrices. Since $T_\theta$ is an interior point of the convex space of qubit channels, it can be represented as a mixture of extremal channels with a smooth $\theta$-dependent probability distribution over such channels. By applying the `classical simulation' argument of \cite{Guta&Rafal}, we conclude that  the QFI $F(\rho^n_\theta)$ grows at most linearly with $n$ and therefore, the estimation rate in the standard metrology setup is $n^{-1}$. Consider now that by monitoring the environment, we know which of the unitaries has been applied on each qubit. Recall that for a rotation family of pure states $|\psi_\theta\rangle= \exp(i\theta G)|\psi\rangle$ the QFI has the expression
\begin{equation}\label{eq.q.fisher.pure}
F_q(|\psi_\theta\rangle) = 4{\rm Var}(G):=  4 (\langle G^2 \rangle - \langle G\rangle^2)
\end{equation}
where $\langle \cdot \rangle$ denotes the expectation with respect to $|\psi\rangle$. If the probe is prepared in the state 
$$
|\Psi^{n}\rangle=\left (|0\rangle^{\otimes n} +|1\rangle^{\otimes n} \right )/\sqrt{2}
$$ 
then $F_{cl}(\theta)=0$ and using \eqref{eq.q.fisher.pure} we find 
$$
F^n_{total} (\theta)= 4 n^2 \lambda_3^2 + 4n( \lambda_3(1-\lambda_3) +\lambda_1 + \lambda_2)
$$
which scales quadratically in $n$.

Before proceeding to the preparation stage, we would like to briefly comment on the physical realizability of our setting. Although it is not our purpose to construct concrete physical models, we point out that `environment monitoring' and continuous time filtering (or quantum trajectories) are well established tools in quantum optics \cite{Gardiner&Zoller}, which been used successfully e.g. for mitigating decoherence \cite{Murch}, speeding up purification \cite{Wiseman&Bouten}, or preparing a target state by means of feedback control \cite{Haroche}. Therefore we believe that the input-output formalism offers a natural framework for continuous time metrology with open systems.

\section{Markov generated probe states} \label{sec.Markov}
\begin{figure}[H]
   \begin{center}  
 \scalebox{0.4}{\includegraphics[width=0.99\textwidth]{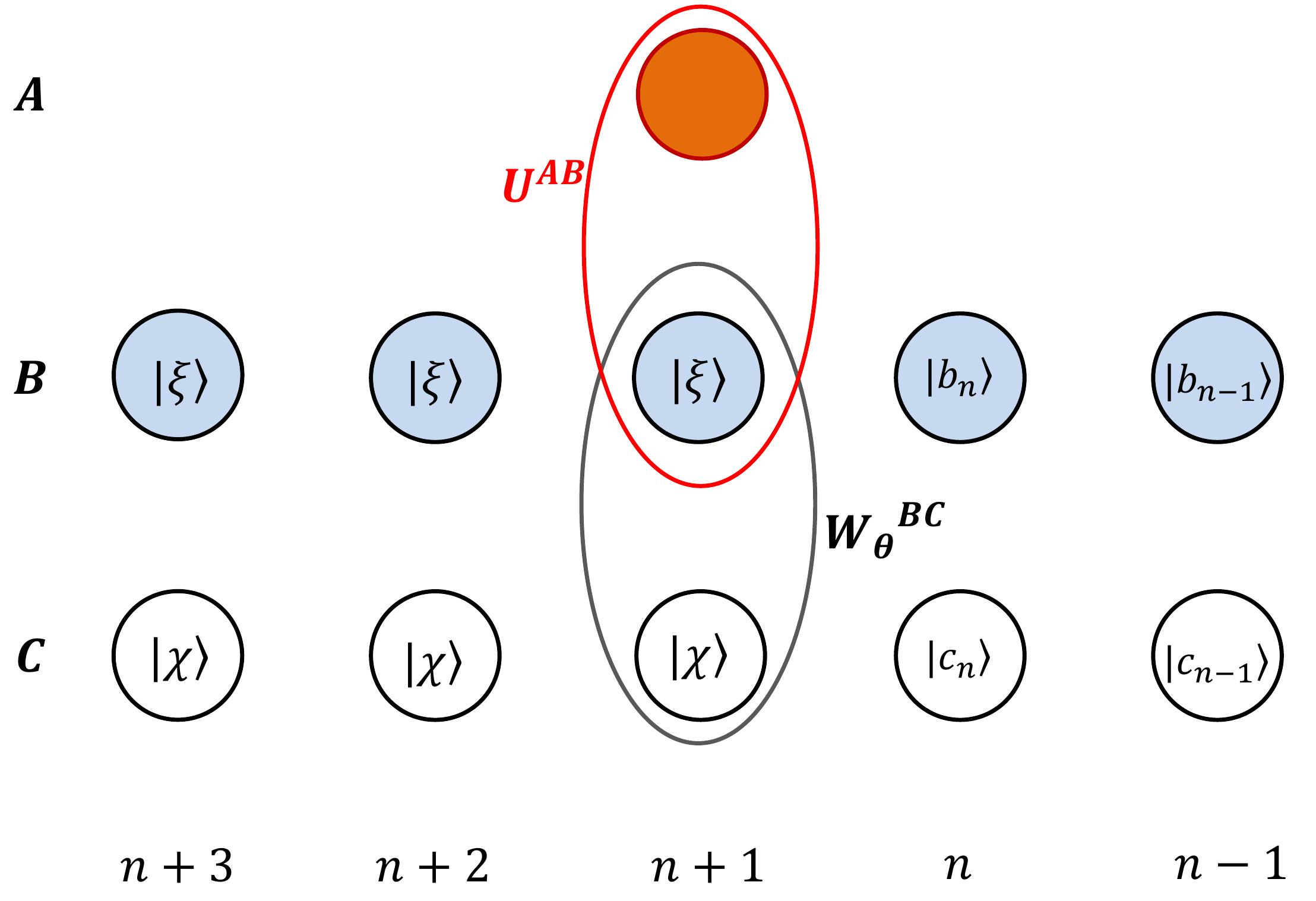}}
   \end{center}
\caption{(Color online) Model of discrete dynamics with Markov generated probe state. 
The memory system $A$ interacts successivly (from right to left) with the probe systems $B$ via the unitary $U^{AB}$. 
The channel $T_\theta$ on $B$ is implemented by unitary coupling of the probe systems with ancillas $C$. }
\label{pic:markov}
\end{figure}

We now introduce the second main ingredient of our analysis: a Markovian mechanism for generating the initial probe state $|\Psi^n\rangle$. This ansatz is partly motivated  by the close relationship to finitely correlated states \cite{Fannes&Nachtergaele&Werner} and matrix product states (MPS) \cite{Wolf&Cirac}, which provide efficient and tractable approximations of complex many-body states \cite{Schollwock}. The preparation stage and subsequent metrology protocol is illustrate in Figure \ref{pic:markov}. The top row represents a `memory system' $A$  which interacts sequentially 
(moving from right to left) with a chain of $n$ identically prepared probe systems (row B), by applying the same unitary $U^{AB}$.  After the interaction, the chain B together with the memory are in the state 
$$ 
|\Psi^n_{AB}\rangle = \sum_{f, \underline{b}}\langle f | K_{\underline{b}}| i\rangle |f\rangle\otimes|\underline{b}\rangle,
$$
where $|i\rangle$ is the initial state of $A$, $\underline{b}= (b_1, \dots , b_n)$  is the index of the product basis for the B row, $K_{\underline{b}}:=K_{b_n}\dots K_{b_1}$, and $K_{b}:=\langle b|  U^{AB} |\xi\rangle$ are the Kraus operators associated to the unitary $U^{AB}$ and the initial state $|\xi\rangle$ of the $B$ systems. 

After the preparation stage, each system undergoes a separate unitary interaction $W_\theta^{BC}$ with an ancilla 
(environment) in row C, prepared initially in state $|\chi\rangle$, as described in the previous section. 
In particular the channel $T_\theta$ and its Kraus operators are given by equation \eqref{eq.ttheta}. By commutativity, the final ABC state is the same irrespective of whether  the unitaries 
$W_\theta^{BC}$ are applied at the end of the preparation stage or each of them is applied immediately after the corresponding $U^{AB}$. With similar notations as above, the joint $ABC$ final state is 
\begin{eqnarray}
&&|\Psi_{ABC}^{n}(\theta)\rangle = \sum_{f, \underline{b},\underline{c}}\langle f| K^{\theta}_{\underline{b},\underline{c}}| i\rangle |f\rangle\otimes|\underline{b}\rangle\otimes|\underline{c}\rangle,
\label{eq:jstate}\\
&=& \sum_{\underline{c}} \left( \mathbf{I}_A\otimes A_{\underline{c}}^{\theta} |\Psi_{AB}^{n}\rangle \right) \otimes |\underline{c}\rangle
= \sum_{\underline{c}} |\tilde{\Psi}_{AB}^n(\theta|\underline{c}) \rangle \otimes | \underline{c}\rangle
\nonumber
\end{eqnarray}
where $|\tilde{\Psi}_{AB}^n(\theta|\underline{c}) \rangle$ is the unnormalised conditional state of $AB$, for a given outcome $\underline{c}$, and $K_{b,c}^{\theta} $ are `extended' Kraus operators. 

The reduced evolution of $A$ is obtained either
by tracing out the $B$ systems in $|\Psi^n_{AB}\rangle$ or the $B$ and $C$ systems in
$|\Psi_{ABC}^{n}\rangle$, and its one step transition operator is 
$$
Z_\theta(\rho) = \sum_{b} K_{b} \rho K_{b}^{\dagger} = 
\sum_{b,c} K_{b,c}^\theta \rho K_{b,c}^{\theta\dagger}.
$$
To summarise, the metrological probe is prepared via the Markov dynamics involving the memory  A and the row B. The channels $T_\theta$ are modelled via the subsequent interaction $W^{BC}$ between the B and the corresponding C systems. We distinguish two scenarios:  the experimentalist has access only to the $B$ systems (which leads to standard rates for non-unitary channels $T_\theta$), or the environment $C$ can be monitored and the collected data can be used to improve the estimation rates.

\vspace{2mm}

From a system identification perspective, this set-up has been investigated in \cite{Guta,Guta&Kiukas},  which show that if the transition operator $Z_\theta$ is \emph{primitive} \cite{Wolf} (memory $A$ converges to a unique stationary state) then the quantum Fisher information of the state $|\Psi^n_{ABC}(\theta)\rangle$ increases linearly with $n$, so $\theta$ can only be estimated with rate $n^{-1}$.
Therefore, from the metrology viewpoint it is interesting to consider models in which the memory $A$ has several invariant subspaces, or `dynamical phases'. In this case, the quantum Fisher information of the full state $|\psi^n_{ABC}\rangle$ may increase as $n^2$ \cite{Guta}, as we will explain below.

\vspace{2mm}

\begin{figure}[H]
   \begin{center}  
 \scalebox{0.45}{\includegraphics[width=0.99\textwidth]{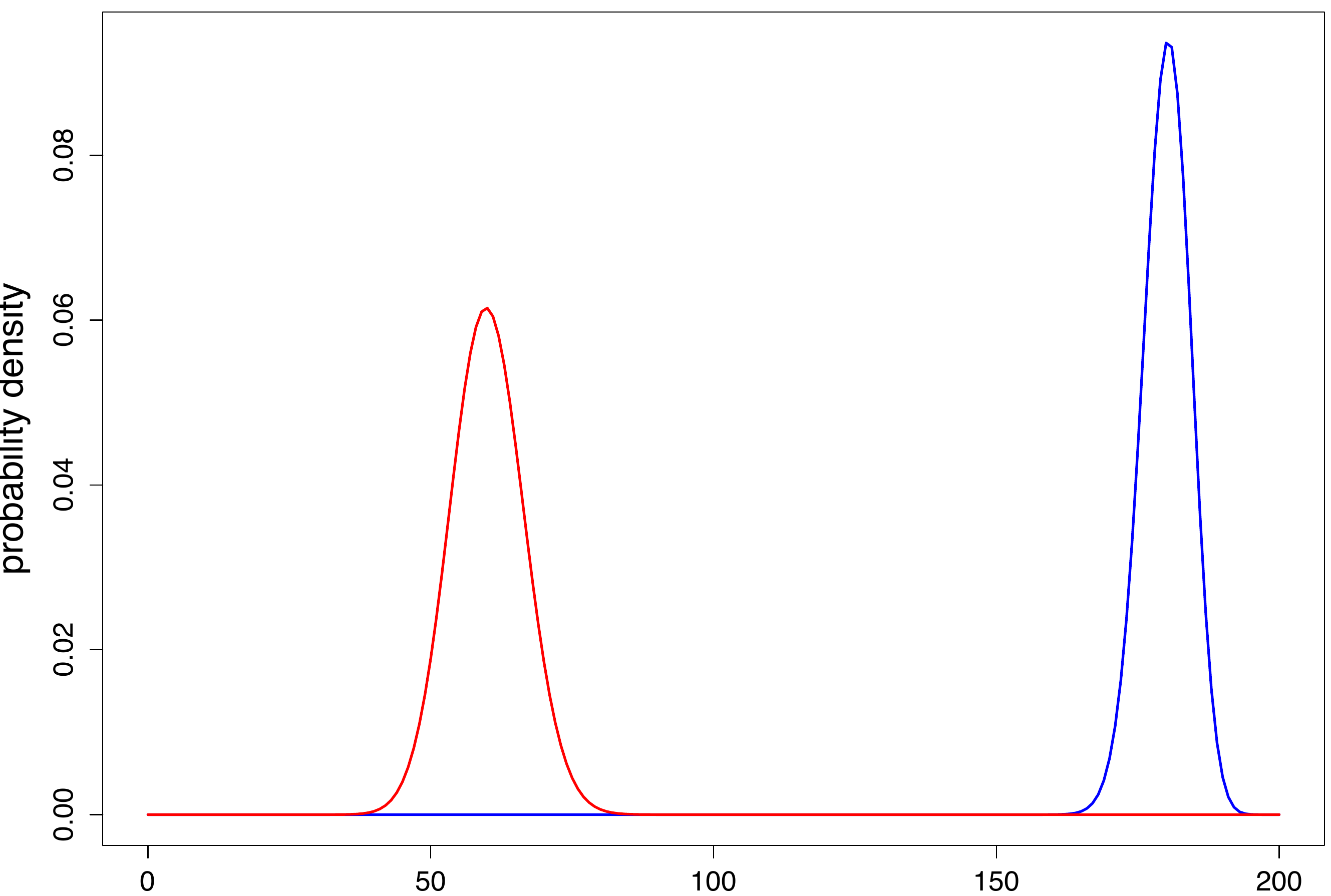}}
   \end{center}
\caption{ (Color online) The distribution of the `total generator' $\bar{G}$ in $|\Psi^n_{ABC}\rangle$ is a mixture of (approximately) Gaussian distributions centred at $ng_0$ and $ng_1$, cf. \eqref{eq.CLT}. When $g_0\neq g_1$ the distance between the two peaks is $n(g_1-g_0)$ and the quantum Fisher information ($F= 4{\rm Var}(\bar{G})$) scales as $n^2$. If phase purification occurs due to `which phase' information leaking to the environment, one of the peaks decays and the variance scales as $n$.
}
\label{fig:binomials}
\end{figure}

For two dynamical phases for instance, the memory space decomposes into orthogonal subspaces 
$$
\mathcal{H}_A= \mathcal{H}_{A}^0\oplus \mathcal{H}_{A}^1,
$$ 
such that the Kraus operators $K_b$ (and similarly for $K_{b,c}$) are block-diagonal with respect to this decomposition 
$$
K_{b}=\langle i | U^{AB} |\xi\rangle= 
\begin{pmatrix}
K_b^0& 0 \\
0 &K_b^1
\end{pmatrix},\quad 
$$
and the restricted evolutions are primitive, with unique stationary states $\rho_{ss}^{0},\rho_{ss}^{1}$. Assuming that the initial state of $A$ is a coherent superposition of states from the two phases, e.g. 
$$
|i\rangle = (|i,0 \rangle +|i,1 \rangle)/\sqrt{2} \in\mathcal{H}_A ,
$$ 
the joint states $|\Psi_{AB}^n\rangle$ and $|\Psi_{ABC}^n(\theta)\rangle$ have a similar decomposition 
$$
|\Psi_{AB}^n\rangle = \frac{1}{\sqrt{2}} 
\left(|\Psi_{AB}^{n,0} \rangle + |\Psi_{AB}^{n,1}\rangle \right)\in \mathcal{H}_A^0 \otimes \mathcal{H}_B^{\otimes n} \oplus
  \mathcal{H}_A^1 \otimes \mathcal{H}_B^{\otimes n}. 
$$
For concreteness we consider a unitary $W^{BC}_\theta$ of the form 
$
W^{BC}_\theta = (U^B_\theta \otimes \mathbf{I}^C) V^{BC} 
$
where $U^B_\theta = \exp(-i\theta G) $ is a phase rotation on the probe system $B$ with generator $G$ and $V^{BC}$ is a fixed unitary describing the interaction with the environment. Since $|\Psi^n_{ABC}\rangle$ is a rotation family, 
$$
|\Psi^n_{ABC} (\theta)\rangle = \exp(- i \theta \bar{G} ) |\Psi^n_{ABC}(0)\rangle ,\quad \bar{G}= \sum_{i=1}^n G^{(i)},
$$
its quantum Fisher information is proportional to the variance of the `total generator' $\bar{G}$. For simplicity, in the sequel we will  identify the total generator with the random variable obtained by measuring $\bar{G}$. Its probability distribution with respect to $|\Psi^n_{ABC} (\theta)\rangle$ is the mixture $(\mathbb{P}^{0} +\mathbb{P}^{1})/2$ of the distributions corresponding to the two phases, computed from the states $|\Psi^{n,0}_{ABC}\rangle$ and $|\Psi^{n,1}_{ABC}\rangle$. Under each 
$\mathbb{P}_0$ and $\mathbb{P}_1$ separately, the following convergence in law to the normal distribution (Central Limit Theorem) holds \cite{Guta}
\begin{equation}\label{eq.CLT}
\frac{1}{\sqrt{n}} (\bar{G} - n g_a) \overset{\mathcal{L}}{\longrightarrow} N(0, V^a), \quad a=0,1.
\end{equation}
for certain means $g_a$ and variances $V_a$. Therefore, if $g_0\neq g_1$ the distribution of $\bar{G}$ with respect to the output state $|\Psi^n_{ABC}(\theta)\rangle$ has variance of the order $n^2$, and we are in the Heisenberg scaling regime, cf. Figure \ref{fig:binomials}. 
We now investigate what happens when the ancillas in row $C$ are measured, as described in our environment monitoring scheme. Note that the measurement data $\underline{c}$ on its own, carries no information about $\theta$, i.e. $F_{cl}(\theta)=0$ since the unitary rotation is applied at the end, and only on the B row. However, as in the toy example of section \ref{sec.monitored}, the results do contribute to a larger quantum Fisher information, by identifying the pure components 
 $|\Psi^n_{AB}(\theta|\underline{c})\rangle$ of the mixed probe state $\rho^n_\theta$. These conditional states have a similar phase decomposition
$$
|\Psi_{AB}^n( \theta|\underline{c} )\rangle = 
\sqrt{p^n_{0}( \underline{c} )} |\Psi_{AB}^{n,0}(\theta|\underline{c}) \rangle + 
\sqrt{p^n_{1}( \underline{c} )} |\Psi_{AB}^{n,1}(\theta|\underline{c}) \rangle
$$
where $p^n_{a}(\underline{c})$ is the probability that $A$ is in phase $a$, given the outcome $\underline{c}$, and $|\Psi_{AB}^{n,a}(\underline{c})\rangle$ is the posterior states corresponding to the initial 
state $|i,a\rangle\in \mathcal{H}_A^a$. 
By the same argument as above, the variance of $\bar{G}$ with respect to the conditional state 
$|\Psi_{AB}^n( \underline{c})\rangle$ increases as $n^2$ \emph{provided} that the `weights' $p^n_{0}(\underline{c}) $ and $p^n_{1}(\underline{c})$ of the two components of the mixture stay away from the extreme values $0,1$. Unfortunately however, this can happen only in special situations as the following argument shows. Let $\mathbb{Q}^n_0(\underline{c})$ and $\mathbb{Q}^n_1(\underline{c})$ be the probability distributions of measurements on the environment corresponding to the two phases. If these distributions are \emph{different} in the stationary regime, the observer can distinguish between them at a certain exponential rate, similarly to the case of discrimination between two coins with different bias. This means that the conditional probability 
$p^n_{0}( \underline{c} )$ will converge (almost surely along any trajectory) either to zero or to one in the limit of large $n$. We call this  phenomenon \emph{phase purification}, in analogy with that of \emph{state} purification for a system monitored through the environment \cite{Maassen}. Essentially, phase purification occurs when the environment learns about the phase of $A$. Therefore, keeping the coherence between the two outputs $ |\Psi_{AB}^{n,0}(\underline{c}) \rangle $ and  
$|\Psi_{AB}^{n,1}(\underline{c}) \rangle$ requires that the environment measurement does not provide any information to distinguish between the two.

In conclusion, total Fisher information $F^n_{total}(\theta)$ scales as $n^2$ if phase purification does not occur, and scales linearly in $n$ otherwise. In the latter case, the long time behaviour is determined by the average of the Fisher informations corresponding to the two phases. However, in the short term the Fisher information may increase quadratically in $n$ until phase purification destroys the coherence between the two phases.

\section{Example of a Heisenberg limited system}\label{sec.example} 
In this section we present a concrete example exhibiting the two behaviours described above, depending on the choice of Markov dynamics.

Inspired by \cite{Jarzyna&Rafal} we consider a minimalistic example with a two dimensional memory whose phases are the basis vectors 
$|0\rangle $ and  $|1\rangle$, and a two dimensional probe unit $B$ with initial state $|\xi\rangle$. The Kraus operators  are of the form
\begin{equation}\label{eq:krausd}
K_{b}=\langle b | U^{AB} |\xi\rangle= 
\begin{pmatrix}
\sqrt{\alpha_b} & 0 \\
0 & \sqrt{\beta_b}
\end{pmatrix},\quad b=0,1,
\end{equation}
with $\alpha_0+\alpha_1=\beta_0+\beta_1=1$. The initial state of $A$ is the superposition $|i\rangle=\frac{1}{\sqrt 2}(|0\rangle_A+|1\rangle_A)$, and the unitary rotation on the probe system B is $U^B_\theta = \exp(i\theta |1\rangle\langle 1|)$. In the absence of noise, the output state is
 $$
|\Psi^n_{AB}(\theta)\rangle = 
\frac{1}{\sqrt{2}}
\left( |0 \rangle \otimes |\psi_0(\theta)\rangle^{\otimes n}
+ 
 |1 \rangle \otimes |\psi_1(\theta)\rangle^{\otimes n}\right)
$$
where 
\begin{eqnarray*}
&&
|\psi_0(\theta)\rangle=
\sqrt \alpha_0 |0\rangle +e^{i\theta}\sqrt \alpha_1 |1 \rangle, \\
&&
|\psi_1(\theta)\rangle=
\sqrt \beta_0 |0\rangle +e^{i\theta}\sqrt \beta_1 |1\rangle.
\end{eqnarray*}
The distribution of the generator $\bar{G}$ with respect to this state is a mixture of two binomial distributions ${\rm Bin}(n, g_0= \alpha_1)$  and ${\rm Bin}(n, g_1= \beta_1)$, and the quantum Fisher information is  
$$
F(|{\Psi}^n_{AB}\rangle)= 
2n(\alpha_0\alpha_1+\beta_0\beta_1)+n^2(\alpha_1-\beta_1)^2.
$$
We add now a noise model given by phase damping in the direction $v$ on the Bloch sphere
$$
\Lambda^v[\rho]= \sum_{c\in \{0,\pm 1\}} A_c \rho A_c^{\dagger}
$$  
with Kraus operators $A_0=\sqrt{p} \mathbf{I}$, $A_{\pm 1}^v=\sqrt{1-p}|v_\pm\rangle\langle v_\pm|$ for $p \in [0,1]$. By \eqref{eq:jstate}, the record  of ancilla measurement outcomes indicate which of the Kraus operators defined $A_j$ acted on each of the probe systems B. Therefore, the unnormalised conditional output state is
\begin{eqnarray*}
\begin{split}
|\tilde{\Psi}^n_{AB}(\theta|\underline c)\rangle =&& 
|0\rangle_A \otimes  U^B_\theta A_{c_n} |\psi_0\rangle \otimes \dots  \otimes U^B_\theta A_{c_1} |\psi_0\rangle \\
&+&
|1\rangle_A \otimes  U^B_\theta A_{c_n} |\psi_1\rangle \otimes \dots \otimes  U^B_\theta A_{c_1} |\psi_1\rangle
\end{split}
\end{eqnarray*}
The probability distribution $\mathbb{Q}^n(\underline{c})=\| \tilde{\Psi}^n_{AB}(\theta|\underline c)\|^2 $ is the mixture 
$(\mathbb{Q}^n_0 + \mathbb{Q}^n_1)/2$ where both components are product measures (independent samples from $\{0,\pm1\}$) with probabilities 
$$
q^{0,1}_0 = p, \quad  q^{0}_{\pm} =(1-p)  \alpha_\pm, \quad q^{1}_{\pm} =(1-p)  \beta_\pm ,
$$
where $\alpha_{\pm}=|\langle v_{\pm}| \psi_0\rangle|^2, \beta_{\pm}=|\langle v_{\pm}| \psi_1\rangle|^2$. In particular, the weights $p^n_a(\underline{c})$ of the two phases and the quantum Fisher information of the conditional state $|\Psi^n_{AB}(\theta|\underline c)\rangle$ depend only on  the total number for each outcome 
$n_{\pm}$ and $n_0$, the latter being the number of systems which have not been affected by the noise. The Fisher information has the following expression
\begin{eqnarray}\label{eq:QFI}
F\left(n_\pm , n_0\right) &=&  (n_+ + n_-) F_\pm +  4 n_0 (p^n_0 \alpha_0\alpha_1 + p^n_1 \beta_0\beta_1)
\nonumber\\
&+& 4 n_0^2 p^n_0 p^n_1 (\alpha_1 - \beta_1)^2
\end{eqnarray}
where $F_\pm= 4 |\langle v_+|1\rangle|^2 |\langle v_-|1\rangle|^2 $, and the phase weights are 
$p^n_1= p^n_1 (n_\pm, n_0) = 1-p^n_0$ with
$$
p^n_{0}= p^n_0(n_\pm, n_0)= 
\alpha_+^{n_+}\alpha_-^{n_-} / (\alpha_+^{n_+}\alpha_-^{n_-}+\beta_+^{n_+}\beta_-^{n_-}).
$$
%
%
\begin{figure}
   \begin{center}  
 \scalebox{0.5}{\includegraphics[width=0.99\textwidth]{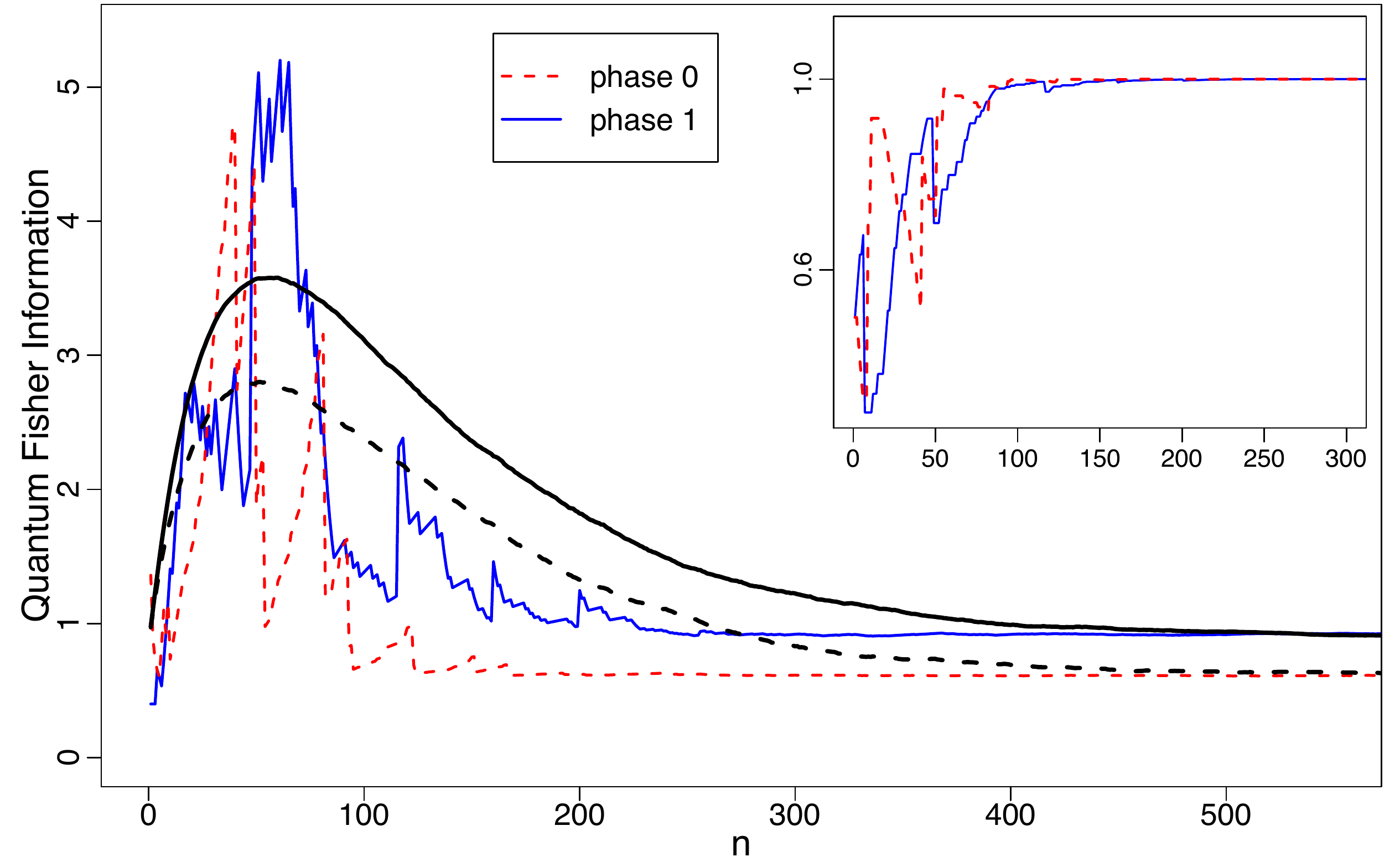}}
   \end{center}
\caption{ (Color online) Phase purification and quantum Fisher information for noise in $x$ direction with parameters $\alpha_0=0.1, \beta_0=0.7$ and $p=0.6$. Main plot: the `scaled' Fisher information $F(|\Psi^n_{AB}(\theta|\underline{c})\rangle) /n$ as function of `time' $n$, for two trajectories with different limiting phases (continuous blue  and  dotted red curves); the corresponding averages are plotted in black. Inset plot:-the weights of the limiting phase $p^n_1(\underline{c})$ and $p^n_0(\underline{c})$ for the two given trajectories converge to $1$ due to phase purification. }
\label{fig:QFIx}
\end{figure}
Since  $n_0\approx pn$, the state has Heisenberg scaling if and only if  the last term in \eqref{eq:QFI} does not converge to zero, i.e. 
the means of $\bar{G}$ in the two phases are different ($\alpha_1\neq \beta_1)$, \emph{and} phase purification does \emph{not} occur ($q^0_\pm = q^1_\pm$).

Now, it is easy to verify that for any noise direction $v$ different from $z$, there exist Kraus operators  $K_b$ such that these conditions are met. In terms of the Bloch sphere, the requirement is that the Bloch vectors of 
$|\psi_0\rangle$ and $|\psi_1\rangle$ lie symmetrically with respect to $v$ such that the (environment) measurement in this direction cannot distinguish between the two states. 

In the special case when $v$ is along the $z$ axis, the means of $G$ in the two phases are equal and therefore the variance scales linearly with $n$. On the other hand, when the environment can distinguish the two phases, either the probabilities $p^n_{0}(\underline{c})$ converges to 1 exponentially with rate equal to the relative entropy $S(q^1 | q^0)$, or 
$p^n_{1}(\underline{c})$ converges to 1 exponentially with rate $S(q^0 | q^1)$. This implies that for \emph{small} $n$ the weights of the two phases are comparable, and  quantum Fisher information \emph{per probe system} $f^n(\underline c)=  \mathbb{E} F_q(|\Psi^n_{AB}(\theta|\underline c)\rangle)/n$ increases linearly with $n$; after that, the exponential decay kicks in and $f^n(\underline c)$ converges to the Fisher information of the corresponding limiting phase. In Figure \ref{fig:QFIx} we illustrate this behaviour with two trajectories having different limiting phases, 
with the phases weights shown in the  inset plot, and the black lines showing the average Fisher information over trajectories converging to either phase. The average of these two is the scaled total information $F^n_{total}(\theta)/n$.   

\section{Conclusions and outlook}
We proposed Markov dynamics as a mechanism for generating probe states for quantum metrology, and showed how the Heisenberg scaling emerges for systems with multiple `dynamical phases', and noiseless channels. Additinally, we modelled noisy channels by coupling the Markov output to `environment' ancillas, and considered the scenario where the environment is monitored to increase the quantum Fisher information of the output. In this setup we found that the survival of the Heisenberg limit depends on whether the environment receives `which phase' information about the memory system. If `phase purification' occurs, the quantum Fisher information of the conditional output state has an initial quadratic scaling, but in the long run the environment wins, and the massive coherent superposition of `output phases' is destroyed 
leading to the standard scaling. However, in a simple example we showed that the Heisenberg scaling is preserved if the Markov dynamics is chosen such that the environment cannot distinguish between the dynamical phases.

These preliminary results open several lines of investigation in the input-output setting with monitored environment, e.g.  finding the `optimal' Markov dynamics and `stopping times' which maximise the constant of the standard scaling, analysing the use of feedback control based on the measurement outcomes. An appropriate framework for answering these questions may be that of 'thermodynamics of trajectories' and `dynamical phase transitions' \cite{Lesanovsky&Garrahan,Lesanovsky&Garrahan&Guta}.

\vspace{1mm}

\emph{Acknowledgements.} We thank Rafal Demkowicz-Dobrzanski, and Katarzyna Macieszczak for useful discussions. This work was supported by the EPSRC grant  EP/J009776/1.

\section*{Appendix}

For reader's convenience we briefly review here some general notions of quantum parameter estimation and quantum 
Fisher information used in the paper. In quantum estimation (or quantum tomography) we are given a system prepared in the state $\rho_\theta$, where $\theta$ is an unknown parameter which for our purposes can be chosen to be one-dimensional, and we would like to estimate $\theta$ by measuring the system and computing an estimator 
$\hat{\theta}= \hat{\theta}(X)$ based on the measurement result $X$. If the map $\theta\mapsto \rho_\theta$ is smooth, then the  following quantum Cram\'{e}r-Rao bound  \cite{Braunstein&Caves} holds for any unbiased estimator 
(i.e.  $\mathbb{E}_\theta (\hat{\theta})=\theta$)
\begin{equation}\label{eq.qcr}
\mathbb{E}[  (\hat{\theta}- \theta)^2 ] \geq  F_q(\rho_\theta)^{-1}.
\end{equation}
The left side is the mean square error (MSE) of $\hat{\theta}$, which is the standard figure of merit for estimation, while the right hand side is the inverse of the quantum Fisher information (QFI), defined below. Note that the left side is an intrinsic property of the quantum statistical model $\theta\mapsto \rho_\theta$, and therefore is a lower bound for the MSE of any unbiased estimator. The QFI is given 
by 
$$
F_q(\rho_\theta)=\rm Tr(\rho_\theta L_\theta^2)
$$ 
where $L_\theta$ is the operator (called symmetric logarithmic derivative) defined by the equation
$$
2\frac{d\rho_\theta}{d\theta}=\{L_\theta,\rho_\theta\}.
$$ 
In particular, if 
$
\rho_\theta= |\psi_\theta\rangle\langle \psi_\theta|
$ is a rotation family with $|\psi_\theta\rangle= \exp(i\theta G)|\psi\rangle$ 
then 
$$F_q(\rho_\theta) = 4{\rm Var}(G):=  4 (\langle G^2 \rangle - \langle G\rangle^2)
$$ 
where $\langle \cdot \rangle$ denotes the expectation with respect to 
$|\psi\rangle$. 

\vspace{2mm}

In general, the quantum Cram\'{e}r-Rao bound \eqref{eq.qcr} may not be achievable, and the restriction to unbiased estimators is not desirable. However, in practice one usually estimates the parameter $\theta$ by performing repeated measurements on an ensemble of $N$ identically prepared systems, and computing an estimator $\hat{\theta}_N$ based on the collected data. In this case, asymptotically optimal estimators (e.g. the maximum likelihood estimator) achieve the Cram\'{e}r-Rao in the sense that  as $N\to \infty$
$$
\sqrt{N}(\hat{\theta}_N - \theta) \overset{\mathcal{L}}{\longrightarrow} N(0, F_q(\rho_\theta)^{-1})
$$
where $\mathcal{L}$ denotes convergence in distribution and $N(\mu, V)$ is the normal distribution of mean $\mu$ and variance $V$. In particular
$$
\lim_{n\to \infty} N \mathbb{E} [  (\hat{\theta}_N- \theta)^2 ] = F(\rho_\theta)^{-1}
$$
and no estimator can improve on this limit for all $\theta$. Therefore the inverse quantum Fisher information is the optimal constant in the large samples scenario which most relevant for practical purposes.

\vspace{2mm}

In the `environment monitoring' setting, we consider the scenario where the parameter dependent state is bipartite 
$\rho_\theta= \rho_\theta^{12}$. Suppose that a projective measurement with outcome $X$ is performed on the second system, and let $\rho^1(\theta|X)$ be the conditional state of the first system, given the outcome. Then the following inequalities hold
$$
F_q(\rho^1_\theta) \leq F_{cl}(X;\theta) +   \sum_x p_\theta(x) F_q(\rho^1(\theta|x)) \leq F_q(\rho_\theta^{12})
$$
where $\rho^1_\theta$ be the reduced state of the first system, $F_{cl}(X;\theta)$ is the classical Fisher information of the measurement result $X$, and $p_\theta(x)$ is the probability of the outcome $X=x$. Obviously the interpretation is that if we measure system 2, the classical information plus the quantum Fisher information of the remaining state is smaller than the full quantum information of both systems and larger than that of the first system alone.

\vspace{2mm}

In the case of quantum metrology the situation is somewhat more complicated, due to the fact that the probe systems are in general correlated rather than independent. Indeed, unitary channels can be estimated with MSE scaling as $N^{-2}$ with the size of the probe ensemble \cite{vGiovannetti&Lloyd}. 
With the notations of the introduction, let us consider an $n$ systems probe ensemble with final state 
$\rho^n_\theta= T^{\otimes n}_\theta (|\Psi^n\rangle \langle \Psi^n|)$ and QFI $F_q(\rho^n_\theta)$. Since the systems may be correlated, the above asymptotic results do not apply automatically. One can however consider a larger ensemble of $N= k\cdot n$ systems consisting of $k$ independent batches of identically prepared sub-ensembles in state $|\Psi^n\rangle$. For large $k$ and given $n$, the quantum Cram\'{e}r-Rao bound is achievable in the sense that
\begin{equation}\label{eq.cr}
N \mathbb{E} [  (\hat{\theta}_N- \theta)^2 ] \approx \frac{n}{F_q(\rho^n_\theta)}.
\end{equation}
By optimising over $|\Psi^n\rangle$ one can in principle minimise the right side and obtain the optimal MSE for a fixed ensemble size $n$. We now distinguish two situations
i) Heisenberg scaling: $F_q(\rho^n_\theta)$ scales as $n^2$ and therefore the left side of \eqref{eq.cr} decreases as 
$n^{-1}$; this indicates that the overall MSE does not scale as $N^{-1}$ but rather as $N^{-2}$. However, the Fisher information theory cannot be used to find the correct asymptotic constant.

ii) Standard scaling: for a large class of noisy channels the quantum Fisher information $F_q(\rho^n_\theta)$ scales as 
$n$ in the sense that the increasing sequence
$$
f_n :=  \sup_{|\Psi^n\rangle} \, \frac{F_q(\rho^n_\theta)}{n}
$$
has a finite limit $f$. By choosing sufficiently large $n$, one can achieve asymptotic MSEs scaling as $1/(f^\prime N)$ for any constant $f^\prime<f$; we conjecture that than by increasing both $k$ and $n$ one can achieve the optimal asymptotic MSE $1/(f N)$. 

\vspace{2mm}

In conclusion, we found that while for Heisenberg scaling the QFI predicts the right scaling of the MSE but not the constant factor, in the standard scaling case the quantum Fisher information predicts the correct asymptotic behaviour of the MSE, including the optimal constant. In Markov generated setting investigated here, the maximum (average conditional) quantum Fisher information is achieved at a finite $n$ and therefore the optimal strategy within this setting is to prepare independent batches Markov correlated states of optimal length.

\end{document}